# How Many Types of Thermodynamical Equilibrium are There: Relation to Information Theory and Holism

*Maria K. Koleva*

**Institute of Catalysis, Bulgarian Academy of Sciences,
1113 Sofia, Bulgaria
e-mail: mkoleva@bas.bg**

Major revision of the thermodynamics is made in order to provide rigorous fundament for functional diversity of holistic type. It turns out that the new approach ensures reproducibility of the information as well.

**INTRODUCTION**

Long-term functional diversity is the most important implement of complex systems for launching their ability to process information. Its simplest form is grounded on the assumption that each node of a model network resides in one of at least two different states each of which represents a logical unit on its input. Then, by means of applying appropriate stimuli, the "logical" landscape of the network can be modified so that to achieve desired logical output. However, while the necessity of functional diversity as fundamental device for processing information is beyond doubt, the existence of more than one stable state needs clarification because of its conflict with the thermodynamics. Indeed, the zero-th law of the thermodynamics asserts that there always exists a single stationary state, called equilibrium, which is global attractor for every initial condition. Therefore, we face the following fundamental dilemma: either the functional diversity is carried out by out-of-equilibrium phenomena or the thermodynamics admits development such that to open the door for more than one type of equilibrium. The elusive point of the first option is the ambiguous role of the kinetics in establishing "near equilibrium" and "out-of-equilibrium" response correspondingly. Indeed, though the demarcation between them goes through the number of admissible long-term stable states (only one long-term stable state is supposed admissible near equilibrium whilst out-of-equilibrium there are more than one admissible long-term stable states) near equilibrium the particularities of the kinetics are supposed irrelevant for the arrival at equilibrium whilst out of equilibrium the same particularities are supposed to govern the realization of the appropriate admissible state. At the same time the dynamics near equilibrium and out of it is assumed identical. In result, these considerations give rise to an evident conundrum, namely: is it feasible to anticipate that the same dynamics generates radically different macroscopic behavior near equilibrium and out of it?! This inconsistency can be viewed as an escalation of the fundamental paradox of the thermodynamics that asserts: dynamics is not a constraint for exerting significant deviations from the equilibrium. Indeed, the assumption that the dynamics is invariant under the reverse of the time opens the door for motion in the backward direction which yields departure from the equilibrium. However, the latter discards the assertion that the equilibrium is global attractor.

The goal of our study is to put the thermodynamics on new grounds and to demonstrate that its revision not only successfully resolves the above paradoxes but gives rise to functional diversity of holistic type as well. In order to delineate better the problem let us elucidate briefly what the functional diversity of holistic type means. The most common definition of the holism reads: a system is holistically organized if its functioning is not reducible to the sum of its constituents. Applied to processing of information, the holism is implemented by the use of any alphabet that has more than two "letters" (logical units). Indeed, though every text in every language is reducible to string of binary units, the opposite is not true, i.e. an arbitrary string of binary units cannot be unambiguously

"translated" into text of a given language if the correspondence between the alphabet and the binary units has not been settled *apriori*. Then, the simplest form of holism is the constitution of an alphabet that comprises more than two logical units ("letters"). The major advantage of the holistic type of processing information is that, apart from the ability for encoding, it initiates the ability of a system to generate information. Indeed, emerging of every new letter makes possible new transcription of the information which gives rise to new "reading" (meaning) of that information. However, from the thermodynamical point of view, every new "letter" suggests new long-term stable state. In turn, the thermodynamics should not only be reconciled with the existence of more than one stable state but it is required to direct the emerging of new ones as well! In addition, the thermodynamics must open the door for reformulating its second law because the generation of new information happens at the expense of entropy decrease which apparently violates the assertion that on the approach to the equilibrium the entropy monotonically increases. Therefore, the major goal of the present study is to ascertain the general route that provides the corresponding entropy changes so that a system not only resides in one of several stable states but is also able to manufacture new ones. An indispensable part of that goal is putting forward a consistent dynamical justification of that route.

## 1. CRISIS OF IDENTITY IN CLOSED SYSTEMS

Turning point of our approach is an overlooked so far inconsistency of the first and the second laws of thermodynamics with the notion of a closed system. Our task is to demonstrate that this discrepancy has far going consequences: it attacks the very idea that a system in equilibrium can be ever isolated from its environment. In the next section we shall illustrate that the proposed by us reformulated laws of thermodynamics remarkably well entangles the resolving of that paradox with our major goal, namely the thermodynamical justification of the functional diversity. We start revealing that conflict by reminding that the bridge between the first and the second law is grounded on the following relation between heat and entropy:

$$dQ = TdS \tag{1}$$

where $Q$ is the heat, $T$ is the temperature and $S$ is the entropy. Together with the first law of the thermodynamics:

$$dU = dQ + dW \tag{2}$$

(1) points out that the entropy approaches its maximum at the expense of exerting transformations "work" $\to$ heat $\to$ entropy; here $U$ stands for internal energy and $W$ for "work". So far it has been taken for granted that this sequence of transformations is to be interpreted as follows: the action of external macroscopic "forces" is transformed to energy dissipation on dynamical level which in turn brings about spatial rearrangement of the system such that the entropy reaches its maximum at equilibrium. At the same time $Q$ and $W$ are considered independent from one another functions. However, it turns out that both assertions are inconsistent and their conflict generates the following insurmountable difficulties:

- Since $Q$ and $W$ are considered independent, the assumption that the entropy reaches maximum at equilibrium imposes constraint only on $Q$ (requiring $dQ \equiv 0$) whilst $dW$ remains arbitrary. In turn, (2) reduces to $dU = dW$ which, however, implies that the internal state of a system, characterized by $U$, becomes explicit function of the environmental circumstances characterized by $W$. Therefore, each time a *closed* system reaches the entropy maximum, its identity has got lost because its behavior is immediately controlled by the environment?! Besides, adding the condition that the thermodynamical equilibrium requires not only maximum of the entropy but maximum of the internal energy as well does not help to improve the situation because generically the maxima of $U$ and $Q$ do not coincide.

- The conditions (1)-(2) are incomplete to provide monotonic increase of the entropy on its approach to equilibrium. Indeed, suppose that a system is subject to constraints that generate some spatial heterogeneity; as an immediate consequence, the entropy becomes spatially dependent. Accordingly, the global entropy maximum, if ever reached, is not anymore a global attractor because some paths go via local minima the escape from which, however, is achieved at the expense of a non-monotonic change of the entropy.

Summarizing, it becomes obvious that the first and the second laws of the conventional thermodynamics suffer severe flaws and must be seriously reconsidered. That is why our top priority is to find out such relation between the identity of a system and the constraints to which it is exposed so that the identity *not* to be immediately controlled by the environment. Besides, as it was already established in the Introduction, the zero-th law also calls for fundamental reconsideration in order to provide rigorous grounds for functional diversity. It will become evident that these problems are counterparts of a successful modification of the thermodynamics. Moreover, the successful reformulation of the all 3 laws of the thermodynamics has to be supplemented by appropriate dynamical and kinetic justification. It is worth noting that the considered above interpretation lacks such justification because the energy dissipation on dynamical level is irreconcilable with any time-reversal dynamics.

## 2. BASIC PRINCIPLES OF THE REVISED THERMODYNAMICS

Our goal is the reformulation of each of the thermodynamical laws so that to at the end to have a self-consistent frame that justifies processing of information of holistic type; besides the new thermodynamics must successfully resolve the considered in the previous section crisis of identity in equilibrium. The first step on this road is the reformulation of the zero-th law. We put forward the assertion that a system is said to be in equilibrium if its macrostate is a *steady* one for any given constraint. Actually, we replace the assumption about equilibrium characterized by a single stationary state with the idea that the domain of admissible constraints reduces to numerous basins of attraction so that the system response within each basin is bounded for providing its long-term steady behavior. One of the major advantages of the new definition is that now the notion of steady state is explicitly related with the functional diversity and processing information. To elucidate this assertion let us point out that the generic function that describes the steady states is every bounded function. Further, since the most general demarcation between information and noise goes through involvement of some causal relations, the bounded functions are classified into two types: the one that certainly involves causal relations (information) comprises all functions that have periodic components and the second one that does not involve any causal relations comprises all functions that have no periodic component. The fundamental reason for that classification is that whist the periodic functions describe a deterministic process that by definition is created by certain causal relations, the "noise" component commences from a stochastic process that is not subject of any causal relation. As a consequence of the above classification, each class of bounded functions serves as logical unit on its input. Thus we give credible basis for the formation of the simplest form of the functional diversity (binary code) but the elucidation how the emerging of holism, i.e. emerging of an "alphabet", happens is still to come. We shall discuss this issue in sec.4 after considering the dynamical justification of the new thermodynamics.

Let us now focus our attention on another aspect of the considered above reformulation. As established in sec.1, the joint action of the first and second law of the conventional thermodynamics yields collapse of the identity of a system in equilibrium. That is why our top priority is to find out such relation between the identity of a system and the constraints to which it is exposed so that the identity *not* to be immediately controlled by the environment. That is why our starting-point is the suggestion that the identity of the system is invariant under minor changes of the constraints. Let us now illustrate how the assumption that the domain of admissible constraints is reduced to numerous basins of attraction so that the system response within each basin is bounded for providing its long-term steady behavior helps to resolve the crisis of the identity. Actually, this assumption replaces the second law with the concept of boundedness which suggests that whatever the behavior of a system is: (i) the distance to the boundaries of every basin is finite; (ii) the rate of exchanging energy/substance with the environment is always bounded. The first aspect is suggested by the assertion that whatever the behavior of the system is, its thresholds of stability are reachable by involving finite amount of energy/substance. The second aspect is a natural expansion of the velocity ansatz which asserts that the rate of transmitting information can not be greater than the speed of the light. The velocity ansatz is crucially important for the reformulation of the first law since we preserve the understanding of the latter as generalized form of energy conservation law. Now we are ready to present the modified first law and to be more easily understandable we shall demonstrate it on the example of an open system:

$$dF = \dot{Q}dt + \dot{W}dt + \mu\dot{N}dt$$
$$dF \equiv 0$$
(3)

where $\mu$ is the chemical potential and $N$ is the number of exchanged species. The novel point is that $F$ does not stand anymore for Gibbs energy but stands for variable called identity energy. Since the latter characterizes the identity of the system, it must be invariant under any changes of the constraints within every given basin. In addition, the required invariance is to be achieved through avoidance of any direct control of the identity energy by the environment. This condition is formally met by the boundedness of each of $\dot{Q}$, $\dot{W}$ and $\dot{N}$ because their mutual dependence does not allow $\dot{Q}$ to become zero at any values of $\dot{N}$ and $\dot{W}$. The required boundedness is automatically guaranteed by the velocity ansatz which imposes exactly boundeness on the rate of exchanging energy/substance with the environment, i.e. it imposes boundedness of each of $\dot{Q}$, $\dot{W}$ and $\dot{N}$. In addition, their sum is permanently constant because it represents energy conservation. In turn, the concept of boundedness certainly makes the invariance of the interaction energy a generic property.

It should be stressed that the reformulated laws are remarkably self-consistent. Indeed, alone the concept of boundedness imposed on (3) not only preserves the identity of the system but explicitly generates the idea of a steady state as equilibrium. Indeed, the boundedness of each of $\dot{Q}$, $\dot{W}$ and $\dot{N}$ and their sum gives rise to bounded functions as functions characterizing the long-term stable behavior (equilibrium); it is easy to check that either each of $\dot{Q}$, $\dot{W}$ and $\dot{N}$ is zero (stationary state) or each of them is a periodic function or a bounded irregular function. Yet, note that the boundeness is a concept available not only at equilibrium but on approach to it as well! Its broadest understanding is that it is a concept which provides the most general conditions for permanent long-term stability of every system.

### 3. DYNAMICAL JUSTIFICATION OF THE REVISED THERMODYNAMICS

A distinctive property of the thermodynamical systems both near equilibrium and out of it is that their macroscopic behavior is described by few variables that have no dynamical analog such as temperature and concentration. Remarkable property of these variables is their insensitivity to the details of the dynamical degrees of freedom. However, the macroscopic behavior near equilibrium and out of it presupposes conflicting dynamical justification, namely: near equilibrium it is assumed that these variables describe irreversible establishing of thermodynamical limit while out-of-equilibrium they are supposed to describe establishing of emergent structures. The conflict is that the same dynamics is supposed to bring about both thermodynamical limit which implies lack of any specific spatio-temporal scale, and emergent structures which involve long-range spatio-temporal correlations. Resolving of the above paradox requires such modification of the dynamics that makes that controversy alias. Moreover, the self-consistency requires plausible explanation of the proposed in the previous subsection modifications of the thermodynamics, namely the boundedness and the extension of the notion of equilibrium from stationary state to steady one. It should be stressed that the conventional time-reversal dynamics cannot explain either of those modifications. Indeed, the time-reversal non-dissipative dynamics opens the door for significant deviations from equilibrium and thus cannot provide the irreversible arrival at equilibrium. On the other hand, the long-range correlations of emergent structures are attacked by the amplification of the local fluctuations [1].

The successful resolving of the paradox is grounded on the following fundamental assumption [2]: we replace the idea of additivity of the interactions with the concept of boundedness. The good reason for the replacement is that whilst the additivity of interactions tacitly allows the interaction energy to become arbitrary on increasing the number of interaction entities and/or their intensity, the boundedness implies that whatever the interaction is its intensity remains bounded. Note that the bounded intensity comes into view as a condition for providing long-term stability.

The major result of our approach developed in [2] is that the dissipation becomes a generic property of any $n-$ body ($n>2$) interaction. Along with the dissipation our approach generates another generic property of the $n-$ body ($n>2$) interactions, namely: the outcome of every many-body interaction is multi-valued function so that only one selection is realized in each interaction. The selection choice is subject of the moment of arrival of the $n-th$ entity to the complex of already interacting $n-1$ entities. The multi-valuedness of the many-body interactions makes the dynamics *not* time-reversible: indeed, since the entities are supposed non-correlated, on turning back time, the selection choice remains random and independent from the selection choice in the forward direction of time. As a result, the dissipation and the multi-valuedness jointly act towards irreversible homogenizing of the system: the larger the local concentration is, the more intensive are those interactions and the dissipation is more intensive. So, the local dissipation brings about local spatial rearrangement that is gradually and irreversibly spread throughout the entire system regardless to the particularities of the dynamics. Eventually, the system reaches state that has no specific spatio-temporal scale and thus can be associated with the maximum of the entropy. The key point is that this result does not bring us back to conventional thermodynamics because the irreversible homogenizing is *not* automatically available for every constraint to which a system is exposed! Whilst it is always available for isolated systems, for the open systems exposed to flux of substance and energy it acts as homogenizer only if the rate of homogenization is faster than the rate of involving substance into the system; otherwise, the many-body interactions destabilize the system because different selections are established at closest points which in turn provokes development of local defects. The elimination of the induced destabilization calls for mechanism that provides long-range coherent behavior so that to keep the system within its thresholds of stability. The task of that mechanism is to impose coherence throughout the entire system insensitively to the particularities of the system. We proved [2] that the implement of the long-range coherent behavior that meets that requirement is non-local feedback between the entities and the collective excitations of the system. Moreover, we proved that it successfully eliminates the induced destabilization at the expense of making the system to exert bounded fluctuations around stationary or periodic state depending on the values and the nature of the external constraints! This result gives credible dynamical justification of our idea that the equilibrium is a not only a stationary state but a steady state that is represented by bounded functions. Thus, it brilliantly justifies the classification of the steady states according to the properties of their periodic components. Besides, it remarkably justifies the idea put forward in the previous section that the domain of admissible constraints is reduced to numerous basins of attraction so that the system response within each basin is bounded for providing its long-term steady behavior. In addition, it gives credible basis for the stability of the emergent structures: their spatio-temporal non-homogeneity is sustained by the balance between the dissipation and the imposing of long-term coherent behavior. Moreover, since both the dissipation and the feedback are insensitive to the details of the dynamics, the emergent structures are also insensitive to them. In the next section we shall demonstrate the crucial role of the emergent structures for providing creation of a new "letter", i.e. for providing holistic type of processing information.

Summarizing, our approach to the dynamics gives plausible basis for the modification of the thermodynamical laws proposed in the previous section. First of all, however, it eliminates the difference between near equilibrium and out-of equilibrium behavior making each concrete system and its constraints subject to different dynamics, i.e. whether it is of homogenizing type or exerts long-range coherence. Indeed, the homogenizing type of dynamics gives rise to conventional thermodynamics while the dynamics that involve long-range coherence gives rise to out-of-equilibrium type of macroscopic behavior. Still, the "switching" between both types of dynamics on manipulating external constraints brings about the unified frame that has been presented in the previous section. Besides, the modified dynamics and the modified thermodynamics are self-consistently entangled by the natural commencing of the macroscopic boundedness from the long-range coherence and proposed ubiquity of the dissipation.

## 4. HOLISM: THERMODYNAMICAL JUSTIFICATION

The demarcation between the information and noise according to whether they are subject of causal relations makes the properties of the power spectrum of every long-term steady state the implement of the functional diversity. Thus, the simplest form of the functional diversity, the binary code, is provided by switching between two long-term stable states the one of which has no discrete band in its power spectrum whilst the other one comprises a discrete band with only one principal frequency. Consequently, the obvious requirement for constituting a new "letter" is emerging of a state that has more than one principal frequency in the discrete spectrum. It is proven in [2] that the binary code is the only one available in spatially homogeneous systems that exerts "long-range coherence" dynamics regardless to how genuine the manipulating of the external constraints is. That is why the creation of every additional long-term stable state that can serve as a new logical unit requires some spatial non-homogeneity. Note that the latter inevitably introduces new periodic component(s) in the discrete spectrum and thus provides the constituting of a new "letter". Therefore, the emerging of every new logical unit must be associated with the emergent structures since they always involve some spatio-temporal non-homogeneity achieved at the expense of permanently sustained balance between the dissipation and the long-range coherence. In turn, the permanently sustained balance provides the long-term stability of the corresponding emergent structure. However, the long-term stability alone is not enough to ensure the non-ambiguous operation of the emergent structure as a logical unit. In addition to its long-term stability, the emergent structure must be reversible, i.e. on releasing the applied external stimulus, it must disappear and it must reappear on applying the same stimulus again. This requirement holds if the modified first law is fulfilled together with the concrete kinetics:

$$dF = \frac{\partial Q}{\partial t}dt + \nabla Q \bullet \frac{d\vec{r}}{dt}dt + \frac{\partial W}{\partial t}dt + \nabla W \bullet \frac{d\vec{r}}{dt}dt + \mu\frac{\partial N}{\partial t}dt + \mu\nabla N \bullet \frac{d\vec{r}}{dt}dt \quad (4)$$

$$dF \equiv 0$$

where $\frac{d\vec{r}}{dt} = \vec{V}$; $\vec{V}$ is the velocity of exchanging substance/energy with the environment and thus it is to be bounded. $\frac{\partial N}{\partial t}$, $\nabla N$, $\frac{\partial W}{\partial t}$, $\nabla W$ are the external stimuli; $\frac{\partial Q}{\partial t}$ and $\nabla Q$ are functionals of the emergent structure created by the kinetics. Actually (4) is not an equation; it is an identity, the fulfillment of which provides reversibility of the emergent structure. It comes to say that not every stimulus generates a logical unit ("letter"): a stimulus that generates logical unit must cause reversible and stable non-linear response. Moreover, in order to produce a new "letter", the stimulus must be non-homogeneous ($\nabla N$ and $\nabla W$ to be non-zero). Then, there is a whole variety of emergent structures ("alphabet") in response to different stimuli. Note, that at the same time there is diffeomorphism between the stimuli and the "letters" – only one "letter" corresponds to a given stimulus.

Yet, there is another type of holism explicitly related to the networks. Its major prerequisite is the spatial diversity: in other words, the system is spatially diversified so that its different parts (nodes) operate independently from one another. Still, it is supposed that each node interacts with its immediate neighborhood. Suppose now that a given node is in one-frequency state. But from the point of view of its neighbors it is a source of a wave. In turn, the wave causes non-autonomous driving of the neighbor nodes and some of them can become in a two-frequency state depending on the intensity of the non-autonomous driving. So this results in local commencing of a new "letter"! Note, however, that random Boolean networks in their to-date form does not take into account this type of holism because their definition requires that whatever the environmental influence is, the set of logical units of each node comprises only $0$ or $1$. i.e. it is always binary.

## 5. BOUNDEDNESS AND ACCURACY OF INFORMATION

A key problem of processing information is outlining the general conditions that provide the separation of the genuine information from the accompanied noise. Here this problem arises from our assumption that the equilibrium is a steady state represented by bounded functions among which are functions that comprise both discrete bands (information) and continuous bands (noise). It is obvious that the reproducibility of the information is met only if the accuracy of extracting information is independent from the statistics of the noise. The ontological aspect of the problem is that the target independence is an indispensable part of preventing the interference between the causal relations that bring about information and the noise. Indeed, note that while the information remains the same on repeating the same external stimuli, the noise realization is ever different. The task of this section is to demonstrate that the problem is successfully resolved in the frame of the modified by us thermodynamics. We shall make evident that the general condition that provides the target independence from the noise statistics is the boundedness of the steady states that symbolize logical units.

Let us consider a bounded function that comprises both a discrete and a continuous band in its power spectrum. We have proved [2,3] that the discrete and the continuous band of every bounded function are additively superimposed; moreover the shape of the continuous band is insensitive to the statistics of the noise. Note that the additivity of the continuous and the discrete band along with the insensitivity of the shape of the continuous band to the noise statistics are the necessary and sufficient conditions for preventing the interference between causality and noise. So, indeed the boundedness is the target general condition that provides the reproducible separation the genuine information from the noise. But now we shall prove more: we shall prove that the only decomposition of the power spectrum compatible with the boundedness is the additivity of the discrete and continuous band. To prove that assertion let suppose the opposite, namely: suppose that the discrete and the continuous band are multiplicatively superimposed:

$$A(t) \propto \int_{f_0}^{\infty} g(f)\exp(ift)df \sum_{l=1}^{\infty} c_l \exp(ilt) \qquad (5)$$

where $A(t)$ is the amplitude of the variations in the time series; $g(f)$ are the Fourier components of the continuous band and $c_l$ are the Fourier components of the discrete band. The purpose for representing the multiplicative superposition of the discrete and continuous band through their Fourier transforms is that it makes apparent the permanent violation of the boundedness of $A(t)$ even when both the noise and the information are bounded. Indeed, it is obvious that there is at least one resonance in every moment regardless to the particularities of the causal relations $c_l$ and the noise $g(f)$. In turn, the resonances break the boundedness of $A(t)$ because they make the local amplitude to tend to infinity. This, however, violates energy conservation law because it implies that finite amount of energy/substance (bounded information and bounded noise) generates concentration of infinite amount of energy/substance (infinitely large amplitude of fluctuations at resonance).

However, it is worth noting that alone the additive superposition of a discrete and a continuous band is not enough to provide the reproducibility of the information. Indeed, the additive superposition reads:

$$A(t) \propto \int_{f_0}^{\infty} g(f)\exp(ift)df + \sum_{l=1}^{\infty} c_l \exp(ilt) \qquad (6)$$

It is obvious that the reproducibility of the genuine information encapsulated in $c_l$ requires insensitivity of $g(f)$ to any noise realization, i.e. to the noise statistics. As we have proved in [2,3], this is possible only for bounded noise; then the continuous band in its power spectrum fits the same shape regardless to the particularities of the noise statistics.

Summarizing, we can conclude that the boundedness is that single general constraint imposed on the states that symbolize logical units which is self-consistent with preventing the interference between causality and noise. The key importance of that result is its immense ubiquity. Indeed, since the boundedness as the constraint that prevents interference between causality and noise is independent from any particularities of the processes that generate time series, it is equally available for systems of different nature ranging from cosmic rays to DNA sequences and financial time series to mention a few.

**CONCLUSIONS**

The goal of the present paper is providing rigorous and self-consistent thermodynamical basis for holistic type of functional diversity. The successful modification of the thermodynamics is essentially grounded on the concept of boundedness viewed as general ontological concept that asserts: each object and subject in the Universe is functionally stable if and only if it comprises and exchanges bounded amount of energy and substance. Remarkable property of our approach is that the concept of boundedness applied simultaneously to the thermodynamics and to the dynamics brilliantly justifies the replacement of the idea that the equilibrium is a single stationary state which is global attractor with the idea that the domain of external constraints reduces to numerous basins of attraction so that the system response within each basin is bounded for providing its long-term steady behavior. Besides, it outlines the conditions at which the conventional thermodynamics is available. One of the major consequences of our approach is that alone the boundedness provides the reproducibility of the information. The enormous effect of that result is its ubiquity; since the boundedness as the constraint that prevents interference between information and noise is independent from any particularities of the processes that generate them, it is equally available for systems of different nature: natural, artificial, social etc.

**REFERENCES**

1. M. K. Koleva, http://arXiv.org/nlin/0601048.
2. M. K. Koleva, http://arXiv.org/physics/0512078
3. M. K. Koleva, http://arXiv.org/cond-mat/0309418